\documentclass[showkeys,amsmath,amssymb,aps,pre,groupedaddress,floatfix,notitlepage]{revtex4-1}

\usepackage{hyperref}
\hypersetup{colorlinks,linkcolor=blue,citecolor=blue,urlcolor=blue}

\usepackage{graphicx}

\begin{document}


\title{\ \ \ \   Jump Processes with Deterministic and Stochastic Controls}

\author{Mark S. Bartlett}
\email[]{Mark.Bartlett@gmail.com}
\affiliation{Department of Civil and Environmental Engineering, Duke University, Durham, NC, USA\\ Department of Civil and Environmental Engineering, Princeton University, Princeton, NJ, USA \\ Stantec, New York, NY, USA
}

\author{Amilcare Porporato}
\email[]{aporpora@princeton.edu}
\affiliation{Department of Civil and Environmental Engineering and Princeton Environmental Institute, Princeton University, Princeton, NJ, USA}

\author{Lamberto Rondoni}
\affiliation{Dipartimento di Scienze Matematiche,
Politecnico di Torino, Corso Duca degli Abruzzi 24, 10129 Torino,
Italy\footnote{Present address: Civil and Environmental Engineering Department and Princeton Environmental Institute,
Princeton University, 59 Olden St, Princeton, NJ 08540, USA}}
\affiliation{INFN, Sezione di Torino, Via P. Giuria 1, 10125 Torino, Italy\\
ORCID:0000-0002-4223-6279}

\date{\today}

\begin{abstract}
We consider the dynamics of a 1D system evolving according to a deterministic drift and randomly forced by two types of jumps processes, one representing an external, uncontrolled forcing and the other one a control that instantaneously resets the system according to specified protocols (either deterministic or stochastic). We develop a general theory, which includes a different formulation of the master equation using antecedent and posterior jump states, and obtain an analytical solution for steady state. The relevance of the theory is illustrated with reference to stochastic irrigation to assess probabilistically crop-failure risk, a problem of interest for environmental geophysics.     \end{abstract}


\maketitle

\section{Introduction}
Control theory applied to complex systems such as networks and small out-of-equilibrium devices has received increasing attention from the physics community \cite{bechhoefer2005feedback,liu2016control,aurell2011optimal}. Less attention has been devoted to systems characterized by random jumps, in spite of the fact that several physical systems experience abrupt, unpredictable transitions that are aptly described as random jumps (see e.g., \cite{belousov2016langevin,bartlett2018state} and references therein). In natural processes jumps are typically external, uncontrolled events; in managed systems, however, jumps also may represent an artificial control that returns the system state to a desired range. Ideally, the control (and representative jumps) should be deterministic, but often a stochastic representation is more appropriate to account for limited and imprecise controls.

Here, for a system with deterministic drift and two jumps processes, representing a natural external process and an artificial stochastic control, we pose a master equation using the recently developed Stratonovich formalism for jump processes \cite{bartlett2018state} along with a novel representation for the jump currents for the control jump process. The latter is especially convenient when the initial and final states of the jump are known (i.e., the set levels for the control). We then solve the master equation for the case of steady state. This class of solutions is general to any control process description; however, the class of solutions is specific to natural processes represented by a nonhomogeneous Poisson jump process forced by inputs with an exponential distribution.

In the second part of the paper we use the developed formalism to represent the soil moisture dynamics forced by random jumps of rainfall and controlled by a state-dependent stochastic irrigation input. We consider different deterministic as well as stochastic scenarios of the irrigation control. The theory can be easily extended to multiple stochastic controls used for redundancy, the details of which will be presented elsewhere. We use the solution also to derive the associated plant water stress that determines crop failure, an important problem in geophysics and environmental engineering \cite{porporato2015ecohydrological}. As shown by the irrigation example, the two different but interchangeable representations of the jump process and the general class of solutions highlight a framework for managing and assessing systems forced by multiple jump processes. Such type of problems also appear in nanoscale systems, in which fluctuations are rarely small and often appear in the form of sudden jumps \cite{muratore2013heat,brandner2015thermodynamics}. Accordingly, several problems of
control at the nanoscale may be tackled with the methods presented here.

\section{Theory}
\subsection{Master Equation}
We consider a system described by a scalar variable $\chi$, evolving in time both deterministically, as described by the drift function $m(\chi,t)$, and randomly, as described by two jump processes. The first one represents an uncontrolled forcing, $\varphi(\chi,t)$, and the second one a controlled forcing, $\check{\varphi}(\chi,t)$, used to reset (via a jump)  the variable from an antecedent state, $\chi^-$, to a posterior state, $\chi^+$, according to a specified protocol. The corresponding Langevin-type equation is
\begin{align}
\frac{d\chi}{dt}=m(\chi,t)+ \varphi(\chi,t)+\check{\varphi}(\chi,t),
\label{eq:SDE}
\end{align}
and thus the term $\varphi(\chi,t)$ is the formal time derivative of a nonhomogeneous Poisson process with arrival rate, $\lambda(\chi,t)$, and state dependent marks, $b(\chi,z)$.
The corresponding master equation for the probability density function (PDF), $p_{\chi}(\chi,t)$, is \begin{equation}\partial_t p_{\chi}(\chi,t)= -\partial_{\chi} \left(J_m(\chi,t)+J_{\varphi}(\chi,t) + J_{\check{\varphi}}(\chi,t)\right).
\label{eq:crossingrates2}
\end{equation}
The drift current is
\begin{align}
&J_m(\chi,t)=m(\chi,t)p_{\chi}(\chi,t),
\label{eq:JmDrift}
\end{align}
while the current for the uncontrolled jump process is conveniently represented using transition probabilities \citep{van2007stochastic,bartlett2018state}, i.e.,
\begin{align}
\partial_{\chi}J_{\varphi}(\chi,t)=p_{\chi}(\chi,t)\int_{-\infty}^{\infty}W(u|\chi,t)du-\int_{-\infty}^{\chi}W_S(\chi|u,t)p_{\chi}(u,t)du,
\label{eq:CurrentJump1}
\end{align}
where $W(\chi|u,t)$ is the transition PDF of jumping away from any prior (antecedent) state $u$ and transitioning to the (posterior) state $\chi$, while $W(u|\chi,t)$ is the transition PDF of jumping from the antecedent state $\chi$ and transitioning to any (posterior) state $u$. Moreover, $\int_{0}^{\infty}W(u|\chi,t)du=\lambda(\chi,t)$, is the frequency of jumps, while the second transition PDF is assumed here to be specific to the Stratonovich interpretation
\begin{align}
W_S(\chi|u,t)&=\frac{\lambda(u,t)p_{z}\left(\eta(\chi)-\eta(u)\right)}{|b(\chi)|},
\label{eq:PDFtranStrat}
\end{align}
where $\eta(\chi)=\int 1/b(\chi)d\chi$, and $p_z(z)$ is the distribution of forcing inputs, $z$. For the latter expression (\ref{eq:PDFtranStrat}), the state $\chi$ is an intermediate value between the states before and after the jump according to the Stratonovich interpretation \citep{van2007stochastic,suweis2011prescription,bartlett2018state}.

For the control jump process, $\check{\varphi}(\chi,t)$, a description of the current based on (\ref{eq:CurrentJump1}) and (\ref{eq:PDFtranStrat}) is not convenient. In fact, it would be preferable to have a description where the initial and final set points of the control are directly specified through their respective distributions. To the best of our knowledge, a representation of this type, as presented in the next section, has not yet been explicitly developed.

\subsection{Characterization by antecedent and posterior PDFs}

Our goal is to express the control jump current explicitly in terms of the distributions of initial and final set points, $\check{p}_{\chi^-}(\chi,t)$ and $\check{p}_{\chi^+}(\chi,t)$, respectively. To this purpose, we start from
\begin{align}
\partial_{\chi}J_{\check{\varphi}}(\chi,t)=p_{\chi}(\chi,t)\int_{-\infty}^{\infty}\check{W}(u|\chi,t)du-\int_{-\infty}^{\chi}\check{W}_S(\chi|u,t)p_{\chi}(u,t)du,
\label{eq:CurrentJumpcurr}
\end{align}
where $\check{W}(u|\chi,t)$ and $\check{W}(\chi|u,t)$ are the transition PDFs of the control process. The first transition PDF, $\check{W}(u|\chi,t)$, represents the probability rate of jumping from the state $\chi$ to any state $u$. It is therefore linked to the Poisson jump frequency, as in (\ref{eq:CurrentJump1}),
\begin{equation}
\label{eq:Wux2}
\int_{-\infty}^{+\infty}\check{W}(u|\chi,t)du=\check{\lambda}(\chi,t).
\end{equation}
Following \citep{bartlett2014excess,daly2007intertime,bartlett2018state}, such a non-homogeneous Poisson arrival rate is linked to the antecedent PDF, $p_{\chi^-}(\chi,t)$, via the average rate of jumping of the control process, $\langle\check{\lambda}(t)\rangle$, as
\begin{align}
p_{\chi^-}(\chi,t)=\frac{\check{\lambda}(\chi,t)p_{\chi}(\chi,t) }{\langle\check{\lambda}(t)\rangle}.
\label{eq:PDFantecedent}
\end{align}

As for the second term on the right hand side of (\ref{eq:CurrentJumpcurr}), the transition PDF can be expressed as the product of the jump frequency and the PDF of the increment in $\chi$,
\begin{align}
\check{W}_{S}(x|u,t)=\check{\lambda}(u,t)\check{p}_{\Delta\chi|u}(\chi|u,t),
 \label{eq:PDFamplitudes}
\end{align}
where $\chi^+=\chi^-+\Delta \chi$, and thus $\check{p}_{\Delta\chi|u}(\chi|u,t)$ may be interpreted as the conditional PDF $ \check{p}_{\chi|u}(\chi|u,t)$. Accordingly, the posterior PDF is defined as
\begin{align}
p_{\chi^+}(\chi,t)=\int_{-\infty}^{+\infty}\check{p}_{\Delta\chi|u}(\chi|u,t)p_{\chi^-}(u,t)du.
\label{eq:PDFposterior1}
\end{align}
Combining (\ref{eq:PDFantecedent}), (\ref{eq:PDFamplitudes}), and (\ref{eq:PDFposterior1}), one obtains
\begin{align}
\int_{-\infty}^{\chi}\check{W}_S(\chi|u,t)p_{\chi}(u,t)du=\langle\check{\lambda}(t)\rangle p_{\chi^+}(\chi,t).
\label{eq:Jux1}
\end{align}

Substituting all the above relationships in (\ref{eq:CurrentJumpcurr}) gives the sought expression for the control jump current
\begin{equation}
\label{eq:CurrentJump2}
\partial_\chi \check{J}_{\varphi}(\chi,t)=-\langle \check{\lambda}(t)\rangle\left(\check{p}_{\chi^+}(\chi,t)-\check{p}_{\chi^-}(\chi,t)\right),
\end{equation}
where the first term is related to the current from jumping from any prior (antecedent) state $u$ and arriving at the (posterior) state $\chi$, while the second component is related to the current from jumping away from the prior state $\chi$ to any posterior state $u$. Because the jump is directly prescribed by this formalism, we do not need to explicitly define the underlying jump process amplitude, $b(\chi, z)$, thus avoiding interpretation issues ({\em e.g.}\ the It\^o-Stratonovich dilemma).
In summary, the master equation (\ref{eq:crossingrates2}) becomes
\begin{equation}\partial_t p_{\chi}(\chi,t)= p_{\chi}(x,t)\int_{-\infty}^{\infty}W(u|\chi,t)du-\int_{-\infty}^{\chi}W_S(\chi|u,t)p_{\chi}(u,t)du
-\langle \check{\lambda}(t)\rangle\left(\check{p}_{\chi^+}(\chi,t)-\check{p}_{\chi^-}(\chi,t)\right).
\label{eq:crossingratesfinal}
\end{equation}

\subsection{General Steady State Solution}

For the master equation (\ref{eq:crossingratesfinal}) in steady state and exponential PDF of the forcing term,
$p_z(z)=\gamma e^{-\gamma z}$, a general solution can be obtained as (see Appendix \ref{sec:MasterSolution} for details)

\begin{align}
p_{\chi}(\chi)=-\frac{e^{-\int\left(\frac{\gamma}{b(\chi)}+
\frac{\lambda\left(\chi\right)}{m\left(\chi\right)}\right)d\chi}}{m\left(\chi\right)}\left(N^{\prime}+
\langle\check{\lambda}\rangle\int h(\chi)e^{\int\left(\frac{\gamma}{b(\chi)}+
\frac{\lambda\left(\chi\right)}{m\left(\chi\right)}\right)d\chi}d\chi\right),
\label{eq:MasterSolutionStep1}
\end{align}
where $N^{\prime}$ is a normalization constant, $b(\chi)=d\chi/dy$ (see Appendix \ref{sec:MasterSolution} Eq. (\ref{eq:ChangeVariable1})) and
\begin{equation}
{h}(\chi)= \frac{\gamma}{b(\chi)}\left(\check{P}_{\chi^-}(\chi)-\check{P}_{\chi^+}(\chi)\right)+\check{p}_{\chi^-}(\chi)-\check{p}_{\chi^+}(\chi).
\label{eq:RenewalDynamics5}
\end{equation}
Note that
$\check{P}_{\chi^-}(\chi)=\int_{-\infty}^{\chi} \check{p}_{\chi^-}(u)du$,
and $\check{P}_{\chi^+}(\chi)=\int_{-\infty}^{\chi} \check{p}_{\chi^+}(u)du$
are the respective cumulative distribution functions (CDFs) of the antecedent and posterior PDFs.

Introducing the potential
\begin{align}
\Phi(\chi)=\int\left(\frac{\gamma }{b(\chi)}+\frac{\lambda(\chi)}{m(\chi)}+\frac{\partial_{\chi}m(\chi)}{m(\chi)}\right)d\chi
= \int\left(\frac{\gamma }{b(\chi)}+\frac{\lambda(\chi)}{m(\chi)} \right) d\chi + \ln \left| m(\chi) \right|
\label{eq:Potential1}
\end{align}
the solution may be written as
\begin{align}
p_{\chi}(\chi)= - e^{-\Phi(\chi)}\left(N^{\prime} + \langle\check{\lambda}\rangle\int h(\chi)e^{\Phi(\chi)}d\chi\right).
\label{eq:MasterSolutionStep2b}
\end{align}
Note that $\langle\check{\lambda}\rangle=0$ in the absence of control processes, in which case
the solution of Eq.~(\ref{eq:MasterSolutionStep2b}) reverts to the one found in Ref.\cite{bartlett2018state}

In lieu of $\langle\check{\lambda}\rangle$, one may consider the crossing frequency, $\upsilon(\xi)$, at the arbitrary level $\chi=\xi$, as defined by

\begin{align}
\upsilon(\xi)=|m(\xi)|p_{\chi}(\xi).
\label{eq:crossingrate1}
\end{align}
For Eq. (\ref{eq:crossingrate1}),  note that the normalization constant of $p_{\chi}(\chi)$ also is a function of the average frequency, $\langle\check{\lambda}\rangle$, i.e.,

\begin{align}
N^{\prime}=\frac{\langle\check{\lambda}\rangle\int_{\chi_{\min}}^{\chi_{\max}} e^{-\Phi(\chi)}\int h(u)e^{\Phi(u)}dud\chi-1}{\int_{\chi_{\min}}^{\chi_{\max}} e^{-\Phi(\chi)}d\chi}.
\label{eq:constNormal2}
\end{align}
Thus, for an assumed value of $\upsilon(\xi)$, one may solve Eq. (\ref{eq:crossingrate1}) for the average frequency, $\langle\check{\lambda}\rangle$.

\begin{figure*}
\includegraphics[width=6 in]{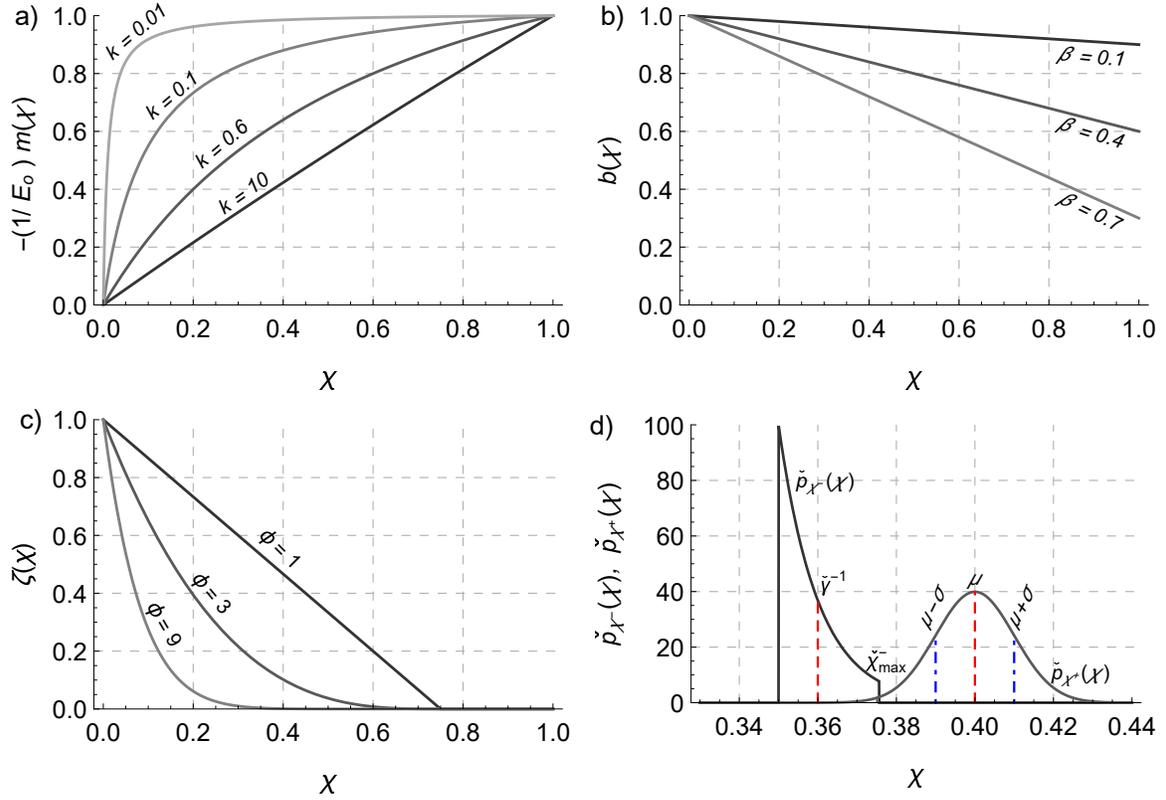}%
\caption{\label{Fig1} Examples of a) the function for evapotranspiration, $m(\chi)$, of Eq. (\ref{eq:DriftIrrigation})
b) the function for water infiltration, $b(\chi)$, of Eq. (\ref{eq:JumpIrrigation}),
c) the function for active water stress, $\zeta(\chi)$, of Eq. (\ref{eq:ActiveWaterStress}), and
d) the posterior PDF, $\check{p}_{\chi^+}(\chi)$, of Eq. (\ref{eq:PDFantIrrigation2}), and antecedent PDF,
$\check{p}_{\chi^-}(\chi)$, of Eq. (\ref{eq:PDFposIrrigation2}), for which we have noted the $\chi$ values
for $\check{\gamma}^{-1}$, $\mu$, $\mu\pm\sigma$, and the value at which the antecedent PDF is truncated,
$\check{\chi}_{\max}^-$. Unless stated otherwise, $\chi_{\max}^+=1$, $\check{\chi}_{\min}=0.35$, $\check{\gamma}=100$, $\sigma=0.01$, $\mu=0.4$, $k=0.5$, $\beta=0.25$, and $f=0.9$.}
\end{figure*}

\section{Application: Soil Moisture Dynamics with Irrigation Control}

\subsection{Soil Moisture and Plant Water Stress}

We apply the previous theory to soil moisture dynamics, a fundamental driver of the terrestrial hydrologic cycles with feedbacks to climate and biogeochemistry (e.g., \cite{rodrigueziturbe2004ecohydrology,katul2007stochastic,porporato2015ecohydrological} and references therein). The soil moisture $\chi$, defined as the relative degree of soil saturation, $0<\chi\leq 1$, jumps because of rainfall infiltration, modeled as a marked Poisson process with constant frequency $\lambda$, and rainfall marks $z$ exponentially distributed with parameter $\gamma=w_0/\alpha$, where $\alpha$ is the mean rainfall depth per event and $w_0$ is the soil water storage capacity. Following \citep{bartlett2014excess}, the infiltration amount, $z\>b(\chi)$, is governed by the function
\begin{align}
b(\chi)=1-\beta\chi,
\label{eq:JumpIrrigation}
\end{align}
where $\beta\chi$ represents runoff to the stream with $\chi$ interpreted in the Stratonovich sense.

During interstorm periods soil moisture decreases mostly because of plant evapotranspiration (ET), modeled as
\begin{align}
m(\chi)=-E_0 \frac{\chi(1+k)}{\chi+k},
\label{eq:DriftIrrigation}
\end{align}
where $E_0$ is potential evapotranspiration and the parameter $k$ adjusts $ET$ with declining soil moisture to  account for different plant water use strategies (Fig. \ref{Fig1}). Eq. (\ref{eq:DriftIrrigation}) accounts for a variety of relationships between evapotranspiration and the soil moisture status \citep{mintz1993global,rodrigueziturbe2004ecohydrology}. As $k\rightarrow 0$, $m(\chi)\rightarrow E_0$; conversely, as $k\rightarrow\infty$, $m(\chi)$ becomes linear (see Fig. \ref{Fig1}).

With these parameterizations, the potential function (\ref{eq:Potential1}) is
\begin{align}
\label{eq:PotentialIrrigation}
&\Phi(\chi)=
\frac{\lambda (k\ln|\chi|+\chi)}{E_0(k+1)}-\frac{\gamma \ln|1-\beta \chi|}{\beta}.
\end{align}
As the soil moisture level declines, plants undergo water stress \citep{rodrigueziturbe2004ecohydrology}, modeled to occur when ET is a fraction $f$ of the potential value, $E_0$, at which point soil moisture is
\begin{align}
\chi^{*}=\frac{f k}{f-k-1},
\label{eq:StresSoil}
\end{align}
which varies with the plant water-use strategy through $k$.
Below this level, $\chi^{*}$, water stress is assumed to increase as
\begin{equation}
\zeta(\chi)=
\left(\frac{\chi^*-\chi}{\chi^*}\right)^{\phi} \quad\quad 0\leq\chi\leq \chi^*,
\label{eq:ActiveWaterStress}
\end{equation}
while it is zero for $\chi>\chi^*$. The parameter $\phi$ accounts for the nonlinear relationship between the soil moisture deficit (from $\chi^*$) and water stress, and $\phi$ reflects plant water use strategies and sensitivity to drought. It is also useful to define
\begin{align}
\langle\zeta^{\prime}\rangle=\frac{1}{P_{\chi}(\chi^*)}\int_0^{1}\zeta \frac{\chi^*\zeta^{\frac{1-\phi}{\phi}}}{\phi}p_{\chi}(\chi^*-\chi^*\zeta^{\frac{1}{\phi}})d\zeta,
\end{align}
an average that only accounts for the continuous part of the PDF $p_{\zeta}(\zeta)$ and thus reflects the average over the typical duration of the stressed condition, $T_{\chi^*}=\frac{P_{\chi}(\chi^*)}{|m(\chi^*)|p_{\chi}(\chi^*)}$. Finally, an effective stress for a growing season of duration $T_s$ may be defined as \citep{rodrigueziturbe2004ecohydrology}
\begin{align}
\theta=\left\{
\begin{array}{l l}
\left(\frac{
\langle\zeta^{\prime}\rangle T_{\chi^*}}{\varpi T_{s}}\right)^{1/\sqrt{n_{\chi^*}}} &\quad\quad \langle\zeta^{\prime}\rangle T_{\chi^*}< \varpi T_{s}\\
\\
1 &\quad\quad \text{otherwise},
\end{array}
\right.
\end{align}
where $n_{\chi^*}=|m(\chi^*)|p_{\chi}(\chi^*)T_s$ is the average number of times the plant enters a stressed condition, and $\varpi$ represents an upper bound for stress prior to permanent plant damage.

\begin{figure*}
\includegraphics[width=5 in]{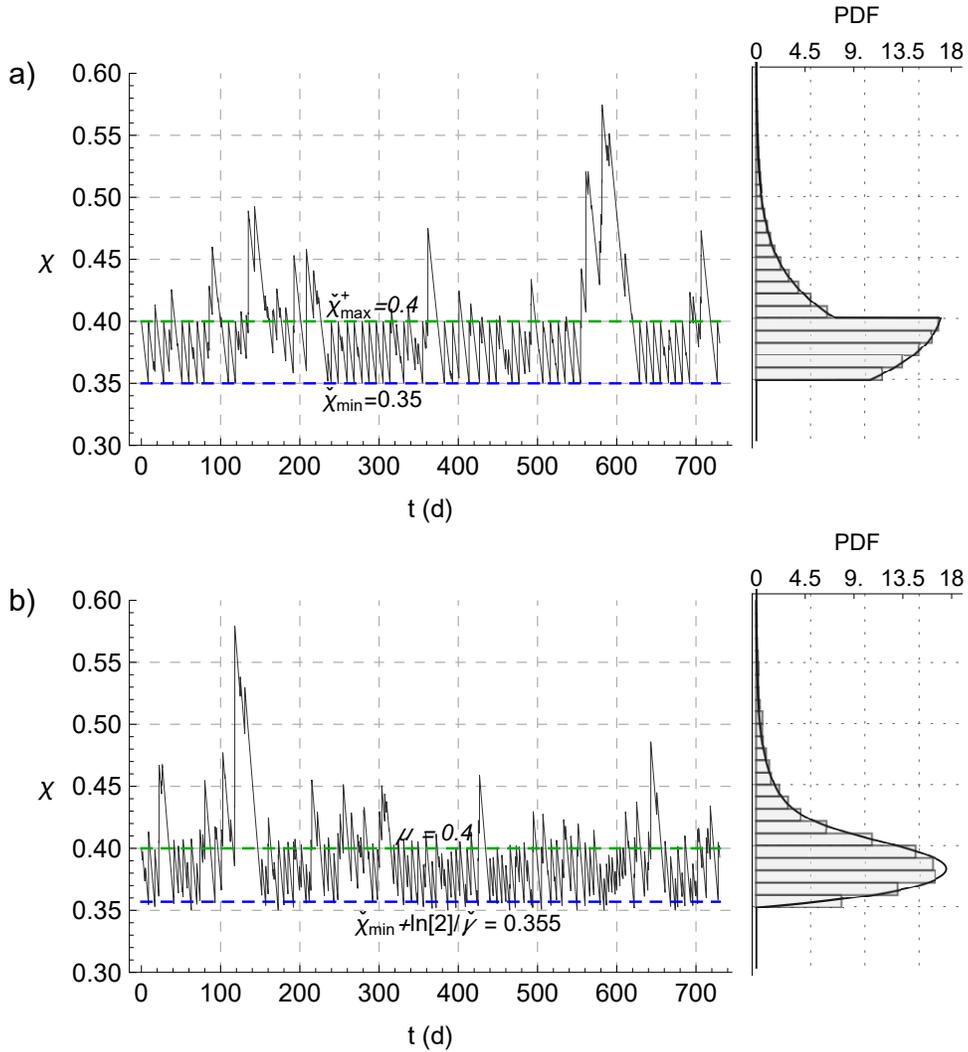}%
\caption{\label{Fig2} Examples of a) deterministic irrigation control and b) stochastic irrigation control soil moisture trajectories, simulation distribution (histogram bars), and the steady-state PDF  (black lines). General parameter values are $\lambda=0.15$ d$^{-1}$, $\alpha=10$ mm, $w_o=450$ mm, $\gamma=w_o/\alpha$, $k=0.5$, $E_0=-4$ mm d$^{-1}$, $\beta=0.25$, $\chi_{\max}^+=1$, and $\check{\chi}_{\min}=0.035$. The stochastic control is based on the antecedent and posterior PDFs of Eqs. (\ref{eq:PDFantIrrigation2}) and (\ref{eq:PDFposIrrigation2}) for which $\sigma=0.01$, $\mu=0.4$, $\check{\gamma}=100$, and $\ln 2/\check{\gamma}$ is the median value of the antecedent PDF. In both cases, the crossing frequency is zero at $\check{\chi}_{\min}$, i.e., $\upsilon(\check{\chi}_{\min})=0$, and $\langle\check{\lambda}\rangle$ may be calculated from Eq. (\ref{eq:crossingrate1}).}
\end{figure*}

\subsection{Irrigation}

Irrigation inputs are introduced to avoid or reduce plant water stress \cite{vico2011rainfed1}. These take place as a control
in the form of instantaneous jumps, a Poisson process with rate obtainable from (\ref{eq:PDFantecedent}),
\begin{align}
\check{\lambda}(\chi)=\langle\check{\lambda}\rangle \frac{\check{p}_{\chi^-}(\chi)}{p_{\chi}(\chi)} ~,
\label{eq:IrrigationFreq}
\end{align}
where $\check{p}_{\chi^-}(\chi)$ represents the distribution of the antecedent before irrigation initiates
(see Fig. \ref{Fig2a}). The applied irrigation water, $\check{z}$, represents a forcing input that increases
soil moisture by the infiltration amount given as
\begin{align}
\Delta\chi=\eta^{-1}\left(\eta(u)+\check{z}\right)-u,
\label{eq:IrrigationAmount}
\end{align}
where following the Stratonovich jump interpretation of the amplitude function, $b(\chi)$, we assume that the infiltration amount decreases as the irrigation input increases based on $\eta(\chi)=\int 1/b(\chi)d\chi$. Following Eq. (\ref{eq:IrrigationAmount}), the irrigation inputs, $\check{z}$, are greater than $\Delta\chi$ because of runoff losses implicit in the function $b(\chi)$ of Eq. (\ref{eq:JumpIrrigation}). In turn, the applied irrigation water of Eq. (\ref{eq:IrrigationAmount}) is part of a joint PDF for the variables governing the irrigation water amount, i.e.,
\begin{align}
\check{p}_{\check{z}}(\check{z},\Delta\chi,u)=\delta\left(\check{z}-(\eta(\Delta\chi+u)-\eta(u))\right)\check{p}_{\Delta\chi|u}(\Delta\chi|u)\check{p}_{\chi^-}(u),
\label{PDFJointIrrigation}
\end{align}
where the Dirac delta function, $\delta(\cdot)$, represents the PDF
$\check{p}_{\check{z}|\Delta\chi u}(\check{z}|\Delta\chi,u)$, i.e.,
the probability density of applied irrigation, $\check{z}$, conditional on the infiltrated irrigation water, $\Delta\chi$, and the antecedent moisture state $u$. The Dirac delta function must be evaluated following the property discussed in Appendix A of \citep{bartlett2014excess}.
By integrating over the PDF of Eq. (\ref{PDFJointIrrigation}), we obtain the average depth of irrigation events,
\begin{align}
\langle \check{z}\rangle = \int_{\Delta\chi_{\min}}^{\Delta\chi_{\max}}\int_{\chi_{\min}^-}^{\chi_{\max}^-}\int_0^{\infty} \check{z}\> \check{p}_{\check{z}}(\check{z},\Delta\chi,u) d\check{z}dud\Delta\chi.
\label{eq:IrrigationAverage}
\end{align}

Based on Eq. (\ref{eq:IrrigationAverage}), the average volume of irrigation water required over a growing season of duration $T_s$ is
\begin{align}
V=w_0\>T_{s}\langle \check{z}\rangle \langle \check{\lambda} \rangle,
\label{eq:VolumeSeason}
\end{align}
using Eq. (\ref{eq:IrrigationAverage}). This volume depends on the soil moisture dynamic described by the steady state PDF $p_{\chi}(\chi)$.

\subsubsection{Deterministic Control}
In an ideal situation of perfectly deterministic control \citep{vico2011rainfed1}, irrigation initiates exactly at the intervention level $\chi=\check{\chi}_{\min}$ and brings the soil moisture to the level $\chi=\check{\chi}_{\max}^+$ (Fig. \ref{Fig2}a), that is
\begin{align}
\label{eq:PDFantIrrigation}
\check{p}_{\chi^-}(\chi)&=\delta(\chi-\check{\chi}_{\min})\\
\check{p}_{\chi^+}(\chi)&=\delta(\chi-\check{\chi}_{\max}^+).
\label{eq:PDFposIrrigation}
\end{align}
These and the respective CDFs (which are right continuous Heaviside step function) define a specific form of the function
$h(\chi)$ of Eq. (\ref{eq:RenewalDynamics5})
The steady state PDF of soil moisture is given by Eq. (\ref{eq:MasterSolutionStep2b}) with substitutions for the
potential of Eq. (\ref{eq:PotentialIrrigation}) and the control process average frequency, $\langle\check{\lambda}\rangle$, and setpoint function, $h(\chi)$,
 of Eq. (\ref{eq:RenewalDynamics5}). As a consequence of the deterministic description of the irrigation control, the solution PDF shows a sharp transition in the probability density at both the initiation level, $\check{\chi}_{\min}$, and the renewal level, $\check{\chi}_{\max}^+$ (Fig. \ref{Fig2}a).

Irrigation occurs as a non-homogeneous marked Poisson process with a state dependent frequency, $\check{\lambda}(\chi)$, defined by Eq. (\ref{eq:IrrigationFreq}) with substitutions for the steady state solution. In conjunction with $\check{\lambda}(\chi)$, the infiltrated irrigation water is distributed as
\begin{align}
\check{p}_{\Delta\chi|u}(\Delta\chi|u)&=\delta(\Delta\chi-(\check{\chi}_{\max}^+-u))
\label{eq:IrrigationInfiltrate}
\end{align}
where irrigation events always increase soil moisture by a deterministic amount of water (per unit area) equal to $w_0(\check{\chi}_{\max}^+-\check{\chi}_{\min})$. The average depth of applied irrigation water is $\langle\check{z}\rangle=\left(\eta(\check{\chi}_{\max}^+)-\eta(\check{\chi}_{\min})\right)$,
which follows from Eqs. (\ref{PDFJointIrrigation}) and (\ref{eq:IrrigationAverage}), both of which are specific to the Stratonovich interpretation of the jump transition. The overall volume of water for a growing season is given by Eq. (\ref{eq:VolumeSeason}).

\begin{figure*}
\includegraphics[width=6 in]{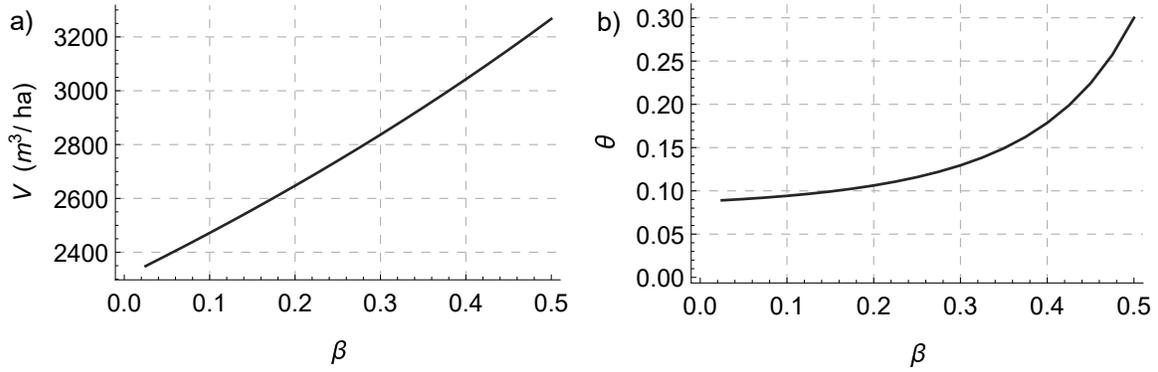}%
\caption{\label{Fig2a} For deterministic irrigation controls over a growing season, $T_s$, and increasing $\beta$, i.e., increasing runoff, a) total volume of applied irrigation water and b) mean effective plant water stress. Parameter values are $\lambda=0.15$ d$^{-1}$, $\alpha=10$ mm, $w_o=450$ mm, $\gamma=w_o/\alpha$, $k=0.5$, $E_0=-4$ mm d$^{-1}$, $\beta=0.25$, $\check{\chi}_{\max}^+=0.4$, $\check{\chi}_{\min}=0.035$, $\varpi=0.5$, $T_s=$120 d, $\phi$=2, and $f=0.9$.}
\end{figure*}

Unlike previous stochastic descriptions of irrigation \citep{vico2010traditional,vico2011rainfed1}, here we account for runoff losses taking place below saturation (through the function $b(\chi)$ of Eq. (\ref{eq:JumpIrrigation})). A low value of $\beta$ could represent an overall efficient management of the agroecosystem that results in both less water losses to runoff, a smaller volume of required irrigation water (Fig. \ref{Fig2a}a), and less water stress. A larger $\beta$ is responsible for a sharp increase in plant water stress (Fig. \ref{Fig2a}b).

\subsubsection{Stochastic Control}

To account of imperfect control, a stochastic irrigation scheme is defined by  $\check{p}_{\chi^-}(\chi)$ and $\check{p}_{\chi^+}(\chi)$, so that the control (on and off) setpoints are now random variables (Fig. \ref{Fig2}b). Here, we represent the antecedent setpoint with an exponential PDF and the posterior setpoint with a normal PDF, both of which are truncated i.e.,
\begin{align}
\label{eq:PDFantIrrigation2}
\check{p}_{\chi^-}(\chi)&=\frac{\check{\gamma} e^{\check{\gamma}(\chi-\check{\chi}_{\min})}}{1-e^{-\check{\gamma} (\check{\chi}_{\max}^- -\check{\chi}_{\min})}}\quad\quad\quad\quad\quad\quad\quad\quad\quad \check{\chi}_{\min} \leq\chi\leq \check{\chi}_{\max}^-\\
\check{p}_{\chi^+}(\chi)&=\frac{2\>e^{-\frac{(\chi-\mu)^2}{2\sigma^2}}}{\sqrt{2\pi}\sigma\>\text{erf}\left[\frac{\check{\chi}_{\min}-\mu}{\sqrt{2}\sigma},\frac{\check{\chi}_{\max}^- +\Delta\chi_{\max}-\mu}{\sqrt{2}\sigma}\right]
}\quad\quad \quad \check{\chi}_{\min}\leq\chi\leq \check{\chi}_{\max}^- +\Delta\chi_{\max}
\label{eq:PDFposIrrigation2}
\end{align}
where erf[$\cdot$] is the error function, and $\mu$ and $\sigma$ respectively control the location and width of the normal PDF, while the exponential PDF scales with the parameter $\check{\gamma}$ (Fig \ref{Fig1}d). The upper bound of the antecedent PDF, $\check{\chi}_{\max}$, is determined from irrigation distribution of soil moisture transitions, for which $\Delta\chi_{\max}$ is the maximum possible transition. We calculate $\langle\check{\lambda}\rangle$ from Eq. (\ref{eq:crossingrate1}) by considering the crossing rate $\upsilon(\chi_{\max})=0$, which corresponds to the case of faultless irrigation for which soil moisture never goes below $\check{\chi}_{\min}$. Figure \ref{Fig2}b shows the soil moisture trajectories and associated steady state PDF of Eq. (\ref{eq:MasterSolutionStep2b}) in this case. The stochastic control diverges from deterministic control according to the respective variances of the PDFs of Eqs. (\ref{eq:PDFantIrrigation2}) and (\ref{eq:PDFposIrrigation2}), i.e., $\check{\gamma}^{-2}$ and $\sigma^2$, respectively. As these variances decrease, the stochastic control approaches the previously discussed deterministic control.

The frequency of irrigation, $\check{\lambda}(\chi)$, approaches infinity as the state variable approaches the lower bound of the setpoint range, i.e., $\lim_{\chi\to \check{\chi}_{\min}^+}\check{\lambda}(\chi)=\infty$, where we assume $\check{\chi}_{\min}$ is approached from the right. Otherwise, if the irrigation control is not faultless, i.e., the crossing frequency is given by $\upsilon(\check{\chi}_{\min})\neq0$,  irrigation failures allow soil moisture levels to decline below $\check{\chi}_{\min}$.

The irrigation control is represented by the frequency $\check{\lambda}(\chi)$ in conjunction with the distribution of soil moisture transitions, $\check{p}_{\Delta\chi}(\Delta\chi$, as defined by Appendix \ref{sec:JumpTransitionPDF} with the $\check{p}_{\chi^+}(\chi,t)$ given by Eq. (\ref{eq:PDFposIrrigation2}), i.e.,
\begin{align}
\check{p}_{\Delta\chi}(\Delta\chi)=\frac{e^{-\frac{(\Delta\chi+\check{\chi}_{\min}-\mu)^2}{2 \sigma^2}} \left(\check{\gamma} \sigma^2-\Delta\chi-\check{\chi}_{\min}+\mu\right)\left(1-e^{-\check{\gamma}(\check{\chi}_{\max}^- -\check{\chi}_{\min})}\right)}{\check{\gamma} \sigma^3\sqrt{2 \pi }\frac{1}{2} \text{erf}\left(\frac{\check{\chi}_{\min}-\mu}{\sqrt{2} \sigma},\frac{\check{\chi}_{\max}^-+\Delta\chi_{\max}-\mu}{\sqrt{2}\sigma}\right)} \quad\quad\quad 0\leq \Delta\chi \leq \Delta\chi_{\max},
\label{eq:IrrigationInfiltrate1}
\end{align}
where the maximum forcing input is
\begin{align}
\Delta\chi_{\max}=\mu-\check{\chi}_{\min}+\check{\gamma}\sigma^2,
\label{eq:zrmax}
\end{align}
which is based on $\check{p}_{\Delta\chi}(\Delta\chi_{\max})=0$. As previously mentioned with Eq. (\ref{eq:PDFantIrrigation2}), the PDF $\check{p}_{\chi^-}(\chi)$ is truncated at an upper bound, $\check{\chi}_{\max}^-$. This upper bound, $\check{\chi}_{\max}^-$, is not known beforehand but is found from the normalization condition $\int_{0}^{\Delta\chi_{\max}}p_{\Delta\chi}(\Delta\chi)d\Delta\chi=1$ (see Appendix \ref{sec:JumpTransitionPDF}). With the antecedent PDF of Eq. (\ref{eq:PDFantIrrigation2}) and the PDF of soil moisture transitions of Eq. (\ref{eq:IrrigationInfiltrate1}), we follow Eq.  (\ref{eq:IrrigationAverage}) and derive the average depth of irrigation water application, i.e.,
\begin{align}
\langle \check{z}\rangle=\int_0^{\Delta\chi_{\max}}\int_{\check{\chi}_{\min}}^{\check{\chi}_{\max}^-} \left(\eta(\Delta\chi+u)-\eta(u)\right)\frac{e^{-\frac{(\Delta\chi+\check{\chi}_{\min}-\mu)^2}{2 \sigma^2}} \left(\check{\gamma} \sigma^2-\Delta\chi-\check{\chi}_{\min}+\mu\right) }{\sigma^3\sqrt{2 \pi }\frac{1}{2} \text{erf}\left(\frac{\check{\chi}_{\min}-\mu}{\sqrt{2} \sigma},\frac{\check{\chi}_{\max}^- +\Delta\chi_{\max}-\mu}{\sqrt{2}\sigma}\right)}e^{-\check{\gamma}(u-\check{\chi}_{\min})}dud\Delta\chi,
\label{eq:IrrigationApplied1}
\end{align}
where $\Delta\chi_{\max}$ is the maximum forcing input of Eq. (\ref{eq:zrmax}). From $\langle \check{z}\rangle$, we calculate the volume of irrigation for a growing season based on Eq. (\ref{eq:VolumeSeason}).

\begin{figure*}
\includegraphics[width=6 in]{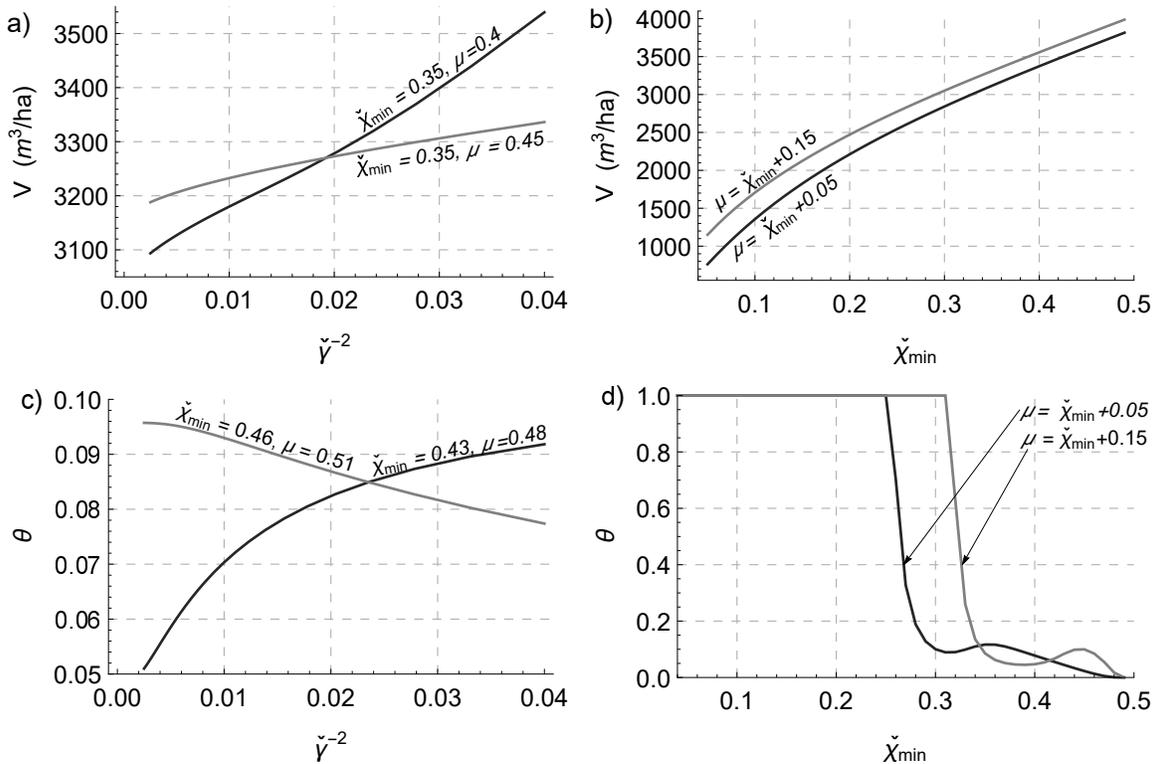}%
\caption{\label{Fig3} In the case of stochastic control, Eqs. (\ref{eq:PDFantIrrigation2}) and
(\ref{eq:PDFposIrrigation2}), examples of a) the volume of irrigation water as a function of the
process noise $\check{\gamma}^{-2}$ where $\sigma^2=0.75\>\check{\gamma}^{-2}$ b) the volume of
irrigation water in terms of the irrigation noise setpoints represented by $\check{\chi}_{\min}$ and $\mu$.
c) the effective water stress, $\Theta$, as a function of the process noise $\check{\gamma}^{-2}$
where $\sigma^2=0.75\>\check{\gamma}^{-2}$, and d) the effective stress as a function of the
irrigation noise setpoints represented by $\check{\chi}_{\min}$ and $\mu$. Unless stated otherwise $\check{\chi}_{\min}=0.35$, $w_o=$
450 mm, $k=$0.12, $E_0=-$4 mm/d, $\beta=0.25$, $\alpha=$10 mm, $\gamma=w_o/\alpha$,
$\lambda=$0.15 d$^{-1}$, $\check{\gamma}=100$, $\mu=0.4$, $\sigma=0.015$, $\chi_{\max}^+=1$, $\varpi=0.5$, $T_s=$120 d, $\phi$=2, and $f=0.9$. In all instances, the crossing frequency is zero at $\check{\chi}_{\min}$, i.e., $\upsilon(\check{\chi}_{\min})=0$, and $\langle\check{\lambda}\rangle$ may be calculated from Eq. (\ref{eq:crossingrate1}).}
\end{figure*}

The volume of required irrigation water varies with both the control noise, as shown by the variances $\check{\gamma}^{-2}$ and $\sigma^2$ (Fig. \ref{Fig1}d), and the range of the control setpoints. The range of the control setpoints approximately is indicated by $\check{\chi}_{\min}$, which represents the lower bound of the antecedent state (the `on' setpoint) and $\mu$, which represents the average renewal state (the `off' setpoint). As the noise increases, the average volume of irrigation water increases more for a smaller versus a larger on/off setpoint range, as roughly indicated by the difference between $\mu$ and $\check{\chi}_{\min}$ (Fig. \ref{Fig3}a). As shown on Fig. \ref{Fig3}a, beyond the noise (variance) of $\check{\gamma}^{-2}\approx 0.02$, the larger range of irrigation on/off setpoints proves more effective in reducing the volume of irrigation water.
Such a behavior occurs because the smaller range of on/off setpoints results in a greater frequency of irrigation as the noise increases. Thus if the farmer or operator cannot precisely control irrigation, it may be more efficient (water wise) to base irrigation on a larger setpoint range with typically large water applications per irrigation event.

Not surprisingly,  as the location of the on/off setpoint range increases,  more irrigation water is required because the soil moisture is being maintained at  values where evapotranspiration is greater (\ref{Fig3}b). The application of irrigation water must be balanced against the benefit it provides in reducing plant water stress, $\theta$, which is considered in terms of the control process noise. Depending on the on/off setpoint range, the control noise (i.e., the variances $\check{\gamma}^{-2}$ and $\sigma$) may either increase or decrease the effective plant stress (Fig. \ref{Fig3}c). Both trends are observed because the effective stress has a minimum value with respect to the location of the on/off setpoints (Fig \ref{Fig3}d). Naturally, the larger setpoint range and thus greater water application per irrigation event reduces the effective stress (black line, Fig \ref{Fig3}d). Notice that the effective stress sharply decreases with the irrigation setpoint range (Fig \ref{Fig3}d). Accordingly, an optimal irrigation control accounts for the inherent noise in the process (Figs. \ref{Fig3}a,c) and provides a setpoint range that both minimizes the volume of irrigation  water (Fig. \ref{Fig3}c) and risk of crop failure as measured by the effective stress (Fig. \ref{Fig3}d).

\section{Conclusions}

We have considered the problem of stochastic jumps with control based on setpoints. A novel formulation of the master equation, based on the mean frequency of jumps of the control process as well as the distributions of the initial and target set-points, permits constructing a master equation where the control appear more transparently than the common formulations with transition probabilities. The steady state solution is expressible in terms of a potential function and a setpoint function, which contains the properties of the control.

We have shown an application to the problem of irrigation, but similar applications can be carried out in other more complex
settings also with multiple controls and redundancies. Such extensions will be presented elsewhere. We also plan to analyze the connections with stochastic thermodynamics of small systems where fluctuations appear as jumps and for which optimal stochastic controls may be especially interesting \cite{muratore2013heat,brandner2015thermodynamics}.

\appendix

\section{Master Equation Solution}
\label{sec:MasterSolution}

For the master equation (\ref{eq:crossingratesfinal}) with exponentially distributed input, $p_z(z)=\gamma e^{-\gamma z}$, we now
derive the steady state solution $p_\chi(\chi)$. To this purpose, we perform a change of variables based on the Stratonovich jump prescription, i.e.:
\begin{align}
\label{eq:ChangeVariable1}
y=\eta(\chi)=\int\frac{1}{b(\chi)}d\chi  ~, ~~ \mbox{ so that } ~~ \chi = \eta^{-1}(y),
\end{align}
so that the corresponding transformed PDF is given by $p_{\chi}(\chi,t)=p_{y}(y,t)\left|\frac{dy}{d\chi}\right|$ and
the master equation (\ref{eq:crossingrates2}) takes the form
\begin{align}
{\partial_t} p_{y}(y,t) =\partial_y(J_m(y,t)+J_{\varphi}(y,t))+\langle\check{\lambda}\rangle(p_{y^+}(y,t)-p_{y^-}(y,t)).
\label{eq:MasterTransformed2}
\end{align}

We then expand Eq. (\ref{eq:MasterTransformed2}) by substituting the probability current terms, i.e.,
\begin{align}
&\frac{\partial}{\partial t}p_y(y,t) = -\frac{\partial}{\partial y}\left[\frac{m\left(\eta^{-1}(y)\right)}{b(\eta^{-1}(y))}p_y(y,t)\right]
- \lambda\left(\eta^{-1}(y),t\right) p_y(y,t)\nonumber\\
&+\int_0^y\lambda(\eta^{-1}(u),t)p_z\left(
y-u\right)p_y(u,t)du\nonumber\\
&+\frac{d\chi}{dy}\delta\left(\eta^{-1}(y)-\chi_{\max}^+\right)\int_{\chi_{\min}}^{\chi_{\max}}\lambda(u,t)\int_{\eta(\chi_{\max}^+)-\eta(u)-(\chi_{\max}^+-u)}^{\infty} p_{z}(1+q-u)p_{\chi}(u,t)dqdu\nonumber\\
&+\langle\check{\lambda}\rangle(\check{p}_{y^+}(y,t)-\check{p}_{y^-}(y,t)).
\label{eq:MasterTransformed3}
\end{align}
Assuming steady state conditions and exponential PDF of $z$, we multiply the equation by the integrating function $e^{\gamma y}$, and differentiate with respect to $y$, i.e.,
\begin{align}
&-e^{\gamma y}\frac{d^2}{d y^2}\left[\frac{m\left(\eta^{-1}(y)\right)}{b(\eta^{-1}(y))}p_y(y)\right]-e^{\gamma y}\gamma \frac{d}{d y}\left[\frac{m\left(\eta^{-1}(y)\right)}{b(\eta^{-1}(y))}p_y(y)\right]-e^{\gamma y}\frac{d}{d y}[\lambda\left(\eta^{-1}(y)\right) p_y(y)]\nonumber\\
&=e^{\gamma y}\langle\check{\lambda}\rangle\left(\gamma \check{p}_{y^-}(y)+\frac{d}{dy}\check{p}_{y^-}(y)-\gamma \check{p}_{y^+}(y)-\frac{d}{dy}\check{p}_{y^+}(y)\right).
\label{eq:MasterTransformed4}
\end{align}
Note that in steady state, the effect of an upper bound, i.e., the fourth term on the r.h.s. of Eq. (\ref{eq:MasterTransformed3}), is accounted for in the normalization constant of the solution PDF \cite{rodrigueziturbe2004ecohydrology}. Dividing both sides of (\ref{eq:MasterTransformed4}) by $e^{\gamma y}$ and integrating,
\begin{align}
&-\frac{d}{d y}\left[\frac{m\left(\eta^{-1}(y)\right)}{b(\eta^{-1}(y))}p_y(y)\right]- \frac{m\left(\eta^{-1}(y)\right)}{b(\eta^{-1}(y))}p_y(y,t)\left(\gamma+\lambda\left(\eta^{-1}(y)\right)\frac{b(\eta^{-1}(y))}{m\left(\eta^{-1}(y)\right)}\right)\nonumber\\
&=\langle\check{\lambda}\rangle\left(\gamma \int \check{p}_{y^-}(y)dy+\check{p}_{y^-}(y)-\gamma \int \check{p}_{y^+}(y)dy-\check{p}_{y^+}(y)\right).
\label{eq:MasterTransformed5}
\end{align}
Multiplying both sides by the integrating function
$$
\exp {\int\left(\gamma+\lambda\left(\eta^{-1}(y)\right)\frac{b(\eta^{-1}(y))}{m\left(\eta^{-1}(y)\right)}\right)dy},
$$
and integrating, after rearranging terms one obtains the desired solution in terms of the transformed variable,
\begin{align}
&p_y(y)=-\frac{b(\eta^{-1}(y))}{m\left(\eta^{-1}(y)\right)}e^{-\int\left(\gamma+\lambda\left(\eta^{-1}(y)\right)\frac{b(\eta^{-1}(y))}{m\left(\eta^{-1}(y)\right)}\right)dy}\nonumber\\
&\left(N+\langle\check{\lambda}\rangle\int\left(\gamma \int \check{p}_{y^-}(y)dy+\check{p}_{y^-}(y)-\gamma \int \check{p}_{y^+}(y)dy-\check{p}_{y^+}(y)\right)e^{\int\left(\gamma+\lambda\left(\eta^{-1}(y)\right)\frac{b(\eta^{-1}(y))}{m\left(\eta^{-1}(y)\right)}\right)dy}dy\right).
\label{eq:MasterTransformed7}
\end{align}
Changing variables again, the solution of Eq.(\ref{eq:MasterTransformed7}) can be given in terms of $\chi$, as in (\ref{eq:MasterSolutionStep1}).

\section{Jump Transition Determined for Exponential Antecedent PDF}
\label{sec:JumpTransitionPDF}
For Eq. (\ref{eq:PDFposterior1}) based on $\check{p}_{\Delta\chi|u}(\Delta\chi,t)=
\check{p}_{\Delta\chi}(\chi-u,t)$, we retrieve the PDF $\check{p}_{\Delta\chi}(\chi-u,t)$ when the antecedent PDF, $\check{p}_{\chi^-}(\chi,t)$, is based on a truncated exponential PDF, i.e.,

\begin{align}
N \check{p}_{\chi^+}(\chi,t) =\int_{0}^{\chi}\check{p}_{\Delta\chi}(\chi-u,t)\frac{\check{\gamma}(t)e^{-\check{\gamma}(t)(u-\check{\chi}_{\min})}}{1-e^{-\check{\gamma}(t)(\check{\chi}_{\max}^--\check{\chi}_{\min})}}\Theta(u-\check{\chi}_{\min})\Theta(\check{\chi}_{\max}^--u)du,
\label{eq:PDFposteriorExp0}
\end{align}
where  $\check{p}_{\chi^-}(u,t)$  now is defined by a truncated exponential distribution, and
$N$ is the normalization constant such that
\begin{align}
N=\frac{1}{\int_{\check{\chi}_{\min}+\Delta\chi_{\min}}^{\check{\chi}_{\max}^-+\Delta\chi_{\max}}\check{p}_{\chi^+}(\chi,t)d\chi},
\label{eq:ConstNormal}
\end{align}
for which $\Delta\chi_{\min}$ and $\Delta\chi_{\max}$ are the respective minimum and maximum jump transitions. We may pose Eq. (\ref{eq:PDFposteriorExp0}) as

\begin{align}
N \check{p}_{\chi^+}(\chi,t)=
\int_{\check{\chi}_{\min}}^{\chi}\check{p}_{\Delta\chi}(\chi-u,t)\frac{\check{\gamma}(t)e^{-\check{\gamma}t)(u-\check{\chi}_{\min})}}{1-e^{-\check{\gamma}(t)(\check{\chi}_{\max}^--\check{\chi}_{\min})}}du,
\label{eq:PDFposteriorExp1}
\end{align}
where the Heaviside step functions of Eq. (\ref{eq:PDFposteriorExp0}) now are implicit in the both the integral limits and in the support of $p_{\chi^+}(\chi,t)$ over the range $\check{\chi}_{\min}+\Delta\chi_{\min} \leq \chi \leq \check{\chi}_{\max}^-+\Delta\chi_{\max}$ as indicated by the normalization constant. Following a substitution for  $\chi=\Delta\chi+\check{\chi}_{\min}$ and then a change of variables based on $u=\Delta\chi+\check{\chi}_{\min}-\upsilon$ \citep{bendat2011random}, Eq. (\ref{eq:PDFposteriorExp1}) becomes

\begin{align}
N \check{p}_{\chi^+}(\Delta\chi+\check{\chi}_{\min},t) = \int_{0}^{\Delta\chi}\check{p}_{\Delta\chi}(\upsilon,t)\frac{\check{\gamma}(t)e^{-\check{\gamma}(t)(\Delta\chi-\upsilon)}}{1-e^{-\check{\gamma}(t)(\check{\chi}_{\max}^--\check{\chi}_{\min})}}d\upsilon.
\label{eq:PDFposteriorExp2}
\end{align}
After multiplying both sides of Eq. (\ref{eq:PDFposteriorExp2}) by $e^{\check{\gamma}(t) \Delta\chi}$ and then differentiating with respect to $\Delta\chi$, we recover

\begin{align}
\check{p}_{\Delta\chi}(\Delta\chi,t)=\frac{ N\left(1-e^{-\check{\gamma}(t)(\check{\chi}_{\max}^--\check{\chi}_{\min})}\right) \partial_{\Delta\chi}\left[\check{p}_{\chi^+}(\Delta\chi+\check{\chi}_{\min},t)e^{\check{\gamma}(t)\Delta\chi}\right]
}{\check{\gamma}(t) e^{\check{\gamma}(t)\Delta\chi}},
\label{eq:PDFforcingPDF}
\end{align}
where $\Delta\chi_{\min} \leq \Delta\chi \leq \Delta\chi_{\max}$ for which the limiting values of $\Delta\chi_{\min}$ and $\Delta\chi_{\max}$  respectively are found by solving $\check{p}_{\Delta\chi}(\Delta\chi_{\min})=0$ and $\check{p}_{\Delta\chi}(\Delta\chi_{\max})=0$.

We also consider the case where the PDF $\check{p}_{\Delta\chi}(\Delta\chi,t)$ and thus jump transition, $\Delta\chi$,
are restricted to positive values, i.e., $\Delta\chi_{\min}=0$. Accordingly, for only positive transitions, we must
restrict the maximum antecedent value, $\check{\chi}_{\max}^-$. The new restricted value of $\check{\chi}_{\max}^-$
must be consistent with the normalization condition,
i.e., $\int_{0}^{\Delta\chi_{\max}}\check{p}_{\Delta\chi}(\Delta\chi)d\Delta\chi=1$. We pose the  normalization condition as

\begin{align}
\frac{\left(1-e^{-\check{\gamma}(t)(\check{\chi}_{\max}^--\check{\chi}_{\min})}\right)}{\int_{\check{\chi}_{\min}}^{\check{\chi}_{\max}^-+\Delta\chi_{\max}}\check{p}_{\chi^+}(\chi,t)d\chi}=
\frac{1}{\int_0^{\Delta\chi_{\max}}\frac{\partial_{\Delta\chi}\left[\check{p}_{\chi^+}(\Delta\chi+\check{\chi}_{\min},t)e^{\gamma(t)\Delta\chi}\right]
}{\check{\gamma}(t) e^{\check{\gamma}(t)\Delta\chi}}d\Delta\chi}
\end{align}
and then find the maximum permissible value of $\check{\chi}_{\max}^-$ for which both sides of the equation are equal.

\begin{acknowledgments}
This work is supported by AFRI Postdoctoral Fellowship program grant no. 2017-67012-26106/project accession Number 1011029;
from the USDA National Institute of Food and Agriculture. This work also was partially funded by the National Science Foundation through grants  EAR-1331846, FESD-1338694, and EAR-1316258.
LR acknowledges partial support from MIUR grant Dipartimenti di Eccellenza 2018-2022, as well as generous hospitality
at the Civil and Environmental Engineering Department and the Princeton Environmental Institute of
Princeton University.
\end{acknowledgments}


\end{document}